\documentclass[twocolumn]{aastex61}

\usepackage[]{natbib}
\usepackage{graphicx}
\usepackage{tabstackengine}
\usepackage{amsmath}
\usepackage[encapsulated]{CJK}
\usepackage{ucs}
\usepackage[utf8x]{inputenc}
\newcommand{\cntext}[1]{\begin{CJK}{UTF8}{gbsn}#1\end{CJK}}

\begin{document}
\title{THE DRAGONFLY NEARBY GALAXIES SURVEY. IV. A Giant Stellar Disk in NGC 2841}

\author{Jielai Zhang (\cntext{张洁莱})}
\affil{Department of Astronomy and Astrophysics, University of Toronto}
\affil{Dunlap Institute for Astronomy and Astrophysics}
\affil{Canadian Institute for Theoretical Astrophysics}
\author{Roberto Abraham}
\affil{Department of Astronomy and Astrophysics, University of Toronto}
\affil{Dunlap Institute for Astronomy and Astrophysics}
\author{Pieter van Dokkum}
\affil{Department of Astronomy, Yale University}
\author{Allison Merritt}
\affil{Max Planck Institute For Astronomy}
\author{Steven Janssens}
\affil{Department of Astronomy and Astrophysics, University of Toronto}

\begin{abstract}
Neutral gas is commonly believed to dominate over stars in the outskirts of galaxies, and investigations of the disk-halo interface are generally considered to be in the domain of radio astronomy. This may simply be a consequence of the fact that deep HI observations typically probe to a lower mass surface density than visible wavelength data. This paper presents low surface brightness optimized visible wavelength observations of the extreme outskirts of the nearby spiral galaxy NGC 2841. We report the discovery of an enormous low-surface brightness stellar disk in this object. When azimuthally averaged, the stellar disk can be traced out to a radius of $\sim$70 kpc (5 $R_{25}$ or 23 inner disk scale lengths). The structure in the stellar disk traces the morphology of HI emission and extended UV emission. Contrary to expectations, the stellar mass surface density does not fall below that of the gas mass surface density at any radius. In fact, at all radii greater than $\sim$20 kpc, the ratio of the stellar to gas mass surface density is a constant 3:1. Beyond $\sim$30 kpc, the low surface brightness stellar disk begins to warp, which may be an indication of a physical connection between the outskirts of the galaxy and infall from the circumgalactic medium. A combination of stellar migration, accretion and in-situ star formation might be responsible for building up the outer stellar disk, but whatever mechanisms formed the outer disk must also explain the constant ratio between stellar and gas mass in the outskirts of this galaxy.

\end{abstract}

\section{Introduction}
The sizes of galaxy disks and the extent to which they have well-defined edges remain poorly understood. Galaxy sizes are often quantified using $R_{25}$, the isophotal radius corresponding to $B=25$ mag arcsec$^{-2}$, but this is an arbitrary choice. In fact, the literature over the last three decades has produced conflicting views regarding whether there is a true physical edge to galactic stellar disks. Early studies seemed to show a truncation in the surface brightness profiles of disks at radii where star formation is no longer possible due to low gas density~\citep{vanderKruitSearle1982}, but more recent investigations have found examples of galaxy disks where the visible wavelength profile is exponential all the way down to the detection threshold~\citep{Bland-Hawthorn2005,vanDokkum2014,vlajic2011}. There is considerable confusion in the literature regarding the relationship between the profile shape and the size of the disk. Most disks fall in one of three classes of surface brightness profile types~\citep{PohlenTrujillo2006,Erwin2008}: Type I (up-bending), Type II (down-bending) and Type III (purely exponential). The existence of Type II disks has been pointed to as evidence for physical truncation in disks. However, while the position of the inflection in the profile can certainly be used to define a physical scale for the disk, this scale may not have any relationship to the ultimate edge of the disk~\citep{Bland-Hawthorn2005,PohlenTrujillo2006}.

The common view in the literature is that the HI disks of galaxies are considerably larger than their stellar disks. This arises from the observation that HI emission extends much further in radius than the starlight detected in deep images~\citep{vanderKruitFreeman2011,Elmegreen2016}. A rationale for this is the possible existence of a minimum gas density threshold for star formation~\citep{FallEfstathiou1980,Kennicutt1989}, although this idea is challenged by the fact that extended UV (XUV) emission is seen in many disks at radii where the disks are known to be globally stable~\citep{Leroy2008}. Studies suggest that large scale instability is decoupled from local instability, and the latter may be all that is required to trigger star formation. For example, a study by~\cite{Dong2008} analyzed the Toomre stability of individual UV clumps in the outer disk of M83. They found that even though the outer disk is globally Toomre stable, individual UV clumps are consistent with being Toomre unstable. These authors also found that the relationship between gas density and the star-formation rate of the clumps follows a local Kennicutt-Schmidt law. In a related investigation,~\cite{Bigiel2010} carried out a combined analysis of the HI and XUV disks of 22 galaxies and found no obvious gas surface density threshold below which star formation is cut off, suggesting that the Kennicutt-Schmidt law extends to arbitrarily low gas surface densities, but with a shallower slope. 

On the basis of these considerations, it is far from clear that we have established the true sizes of galactic disks at any wavelength. Absent clear evidence for a physical truncation, the `size' of a given disk depends mainly on the sensitivity of the observations. This basic fact applies to both the radio and the visible wavelength observations, and relative size comparisons which do not account for the sensitivity of the observations can be rather misleading. For example, it is commonly seen that the gas in galaxies extends much further in single dish observations than it does in interferometric observations, because single dish observations probe down to lower column densities~\citep{Koribalski2016}. At visible wavelengths, the faintest surface brightness probed by observations has been stalled at $\sim 29.5$ mag arcsec$^{-2}$ for several decades~\citep{Abraham2016book}, with this surface brightness `floor' set by systematic errors~\citep{Slater2009}. 

The Dragonfly Telephoto Array (Dragonfly for short) addresses some of these systematic errors and is optimized for low surface brightness observations; see~\cite{Abraham2014} for more details. Dragonfly has demonstrated the capability to  routinely reach $\sim$32 mag arcsec$^{-2}$ in azimuthally averaged profiles~\citep{vanDokkum2014,Merritt2016a}.

In this paper, we present ultra-deep visible wavelength observations taken with Dragonfly of the spiral galaxy NGC 2841. This galaxy is a particularly clean example of XUV emission in an isolated environment~\citep{Afanasiev1999}. It is notable for being the archetype for the flocculent class of spiral galaxies. The disk is globally Toomre-stable~\citep{Leroy2008} and it shows no evidence for grand design structure, although near-infrared observations do show some long dark spiral features in its interior~\citep{Block1996}. Our aim is to determine if the stellar disk, as traced by visible wavelength light, extends at least as far as the neutral gas mapped by the THINGS survey~\citep{Walter2008}, and the XUV emission mapped by GALEX~\citep{Thilker2007}. Instead of just comparing sizes of disks in different wavelengths, we will compare mass surface densities up to the sensitivity limit of the respective data sets. Throughout the paper, we assume the distance to NGC 2841 is 14.1 Mpc~\citep{Leroy2008}.

In \S\ref{sec:data} we describe our observations and the specialized reduction techniques we have adopted in order to obtain deep profiles with careful control of systematic errors. Our results are presented in \S\ref{sec:results}, and our findings are discussed in \S\ref{sec:discussion}. 

\section{Observations and Data Reduction}
\label{sec:data}

Broadband images of NGC 2841 were obtained between 2013 and 2016 using Dragonfly as part of the Dragonfly Nearby Galaxy Survey~\citep{Merritt2016a}. Dragonfly is comprised of multiple lenses with their pointing offset from one another by a few arcminutes. Between 2013 and 2016, the number of lens and camera subsystems on the Array increased from 8 to 24. A total of 3351 ten-minute exposure images of NGC 2841 were obtained in Sloan \textit{g} and \textit{r} bands, distributed over the multiple cameras. Sky flats were taken daily at twilight and dawn. Data reduction was carried out using the Dragonfly Pipeline, full details for which can be found in~\cite{Zhang2018}. The full-width at half maximum (FWHM) of the final combined NGC 2841 image is 7 arcseconds.  

The ultimate limiting factor in low surface brightness observations of nearby galaxies is the wide-angle point-spread function (PSF;~\citealt{Slater2009, Abraham2014, Sandin2015}). The largest-scale component of the PSF is the so-called `aureole'~\citep{Racine1996, King1971}. In conventional telescopes, the aureole is dominated by scattered light from internal optical components ~\citep{Bernstein2007}. An important point emphasized in~\cite{Zhang2018} is that this stellar aureole varies on a timescale of minutes, and so its origin is most likely atmospheric.~\cite{AtmIceCrystals2013} suggest that high-atmosphere aerosols (mainly ice crystals) are the culprit. An efficient way to detect the existence of atmospheric conditions which result in prominent stellar aureoles is to monitor the photometric zeropoints of individual exposures and identify those with deviations from the nominal zeropoint for a given camera at a given air mass. In our analysis of data from NGC 2841, exposures with a photometric zeropoint deviant from the nominal zeropoint by more than $\sim$0.1 mag were excluded from the final combined image. Out of the 3351 exposures obtained, 1034 were used. Most of the exposures excluded were taken in obviously marginal weather conditions (e.g., thin clouds). However, $\sim$25\% of the exposures were identified as having wider-than-normal wide-angle PSFs only by using the procedure of monitoring the zeropoint values of the exposures.

Sky subtraction was done in two passes. In the first pass, a sky model was fit to the \textsc{SExtractor}~\citep{SExtractor} background map for each image and subtracted. Sky-subtracted frames were then used to create an average combined image (including both \textit{g} and \textit{r}-band data). \textsc{SExtractor} was run on this average combined image to produce a segmentation map. A mask was created by growing the segmentation map with settings to capture sources all the way out to their low surface brightness outer edges. In the second pass, sky models were fit to the \textsc{SExtractor} background map of non-sky-subtracted images again, but this time the mask was input into \textsc{SExtractor} for the creation of the background map. This ensures we do not over subtract the sky by fitting a sky model to ultra-faint galaxy light. After this careful sky subtraction procedure, there were no residual large-scale gradients visible to the eye in an image where non-sky pixels were masked. In order to measure any residual large-scale gradients not obvious by visual inspection, a third order polynomial was fit to a masked image. The peak to peak range of sky background model values are ~0.05\% the sky value for both the g and r-band images. This was measured on a 57 by 70 arcminute image of NGC 2841, with the long edge aligned in the north-south direction. It is important to note, however, that the regions responsible for the 0.05\% variation in sky are all on the edges of the image, and do not overlap with NGC 2841. 

The sky background and its error was determined by measuring the flux in elliptical annuli placed around NGC 2841 in each of the \textit{g} and \textit{r}-band Dragonfly images. Randomly placed elliptical annuli were used to sample the sky because the error in the sky value depends on the scale over which the sky is measured. For example, the variance in a sample of 10x10 pixel sky boxes will be different to a sample of 100x100 pixel sky boxes. The outer-most data points in the surface brightness profile are the most sensitive to the accuracy of the sky background measurement, therefore, that is the scale on which it is critical to know the sky background error. The shape of the sky sampling “box” was chosen to be an elliptical annulus because that most resembles the shape within which we have to determine the sky for the surface brightness profile. 1000 elliptical annuli were placed randomly in a 30 by 30 arcminute region around NGC 2841. The ellipticity and position angle of the elliptical annuli was fixed to be that of the largest surface brightness profile isophote. The annuli sizes were allowed to randomly vary but not below the size of the largest surface brightness profile isophote. In order to sample the local sky background values, the elliptical annuli were not allowed outside of a 30 by 30 arcminute region around NGC 2841. A mask was created so that no light from any sources was included in the determination of the sky level. First, the segmentation map produced by \textsc{SExtractor}~\citep{SExtractor} was grown to include the faint outer extents of the sources. For the brightest stars, as well as NGC 2841, the mask was then grown further until no light was visible from these sources using the histogram stretch option in SAOImage DS9. The average and standard deviation of all the sky value measurements in the 1000 elliptical annuli were used to define the sky value and the error on the sky value, respectively. The mean sky surface brightnesses in g and r band were 21.1 mag arcsec$^{-2}$ in g-band and 20.2 mag arcsec$^{-2}$ in r band. The percentage errors on these sky levels were 0.01\% and 0.007\%, corresponding to limiting surface brightness levels of 30.9 and 30.6 mag arcsec$^{-2}$ for the g and r-band images, respectively. The error in the sky value is the dominant source of uncertainty in the surface brightness profile at large radii. 

The HI map used in comparisons below was taken from The HI Nearby Galaxy Survey (THINGS), made using data from the NRAO Very Large Array (VLA)~\citep{Walter2008}. We obtained a far UV (FUV) map of NGC 2841  from the Galaxy Evolution Explorer (GALEX) Nearby Galaxies Survey~\citep{GildePaz2007} using the Detailed Anatomy of Galaxies (DAGAL) image repository~\citep{DAGAL2015}.

\section{Analysis and Results}
\label{sec:results}
\subsection{Size of the stellar disk}

\begin{figure*}[!htbp]
\centering
\includegraphics[width=0.8\textwidth]{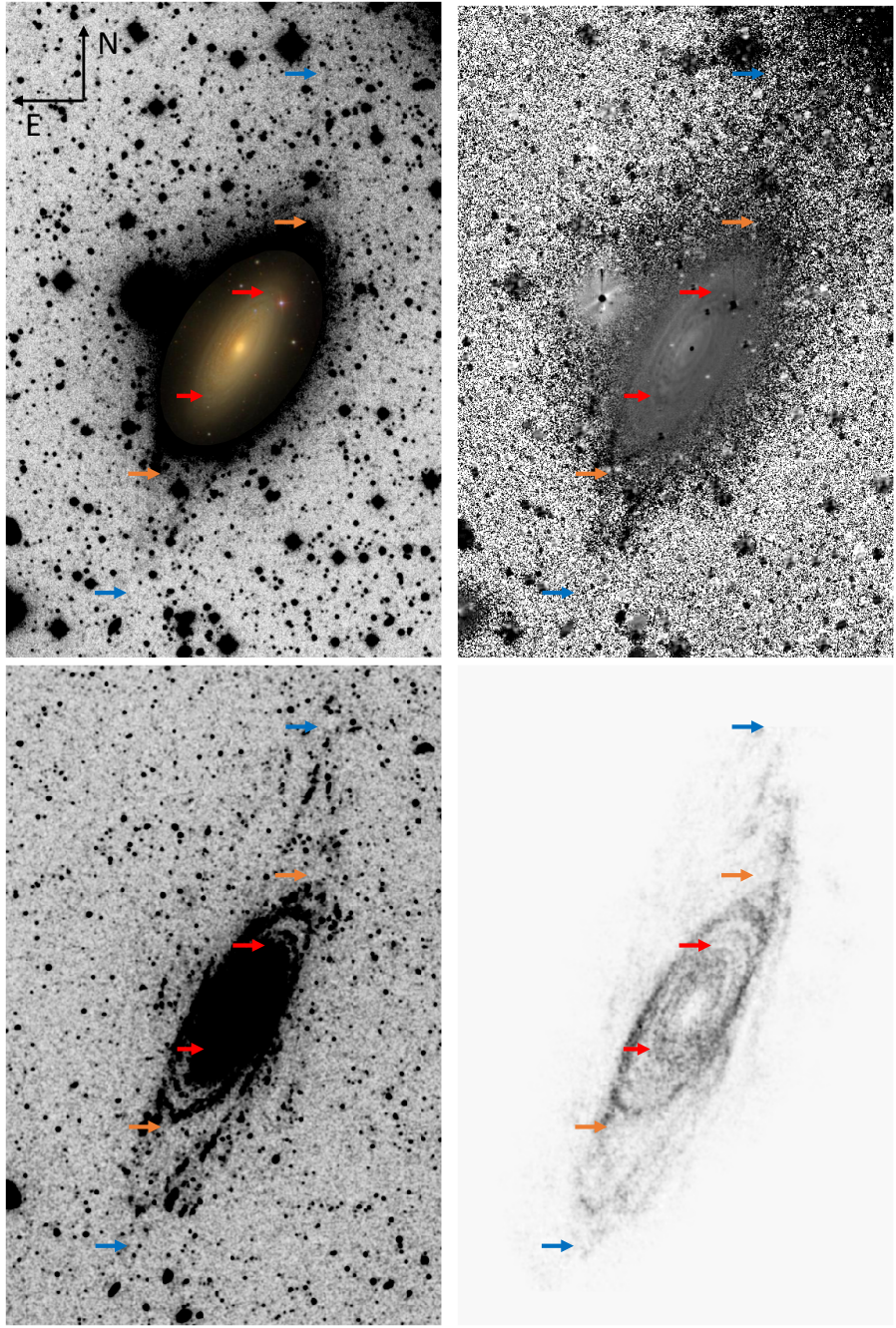}
\caption{A panchromatic view of NGC 2841. Top left: Dragonfly g-band image. The color image embedded in the center was taken from the Sloan Digital Sky Survey data release 14 skyserver website~\citep{SDSSDR14}. Top right: Dragonfly (\textit{g}-\textit{r}) color image. Bottom left: GALEX FUV image. Bottom right: THINGS HI image. The arrows (red, orange, blue) in each image mark the following radii along the galaxy: $R_{25}$ (14.2 kpc), 30 kpc, 60 kpc. The UV, HI gas and visible wavelength emission are traced out to similar radii in the disk of NGC 2841, with the peaks of the visible wavelength disk corresponding to the peaks of the HI and UV emission.}
\label{fig:image}
\end{figure*}

Our images of NGC 2841 in \textit{g}-band and (\textit{g}-\textit{r}) color are shown in Figure~\ref{fig:image}, together with maps of FUV and HI emission. Figure~\ref{fig:image}'s \textit{g}-band image shows a giant disk extending as far as the HI and XUV emission. This disk is visible out to $\sim$60 kpc radius, which is $\sim$4 $R_{25}$ ($R_{25}=14.2$ kpc). To guide the eye, three pairs of arrows are marked in Figure~\ref{fig:image}, colored red, orange, and blue, corresponding to radii of 14.2 kpc ($R_{25}$), 30 kpc, 60 kpc respectively. The orange arrows (30 kpc) mark the edge of the well-studied inner disk of this commonly-observed galaxy~\citep{Block1996,Leroy2008,Silchenko2000,Afanasiev1999}. The disk beyond $\sim$30 kpc appears warped in all three wavelengths. Interestingly, there may be two distinct warped disks, which is most obvious in the (\textit{g}-\textit{r}) color and HI images. The peaks of the HI disk correspond to the peaks seen in the XUV and the visible wavelength data, which suggests that, at large radii, stars in this galaxy are mainly in a disk and are not part of a stellar halo.

\begin{figure}[!htbp]
\centering
\includegraphics[width=0.52\textwidth]{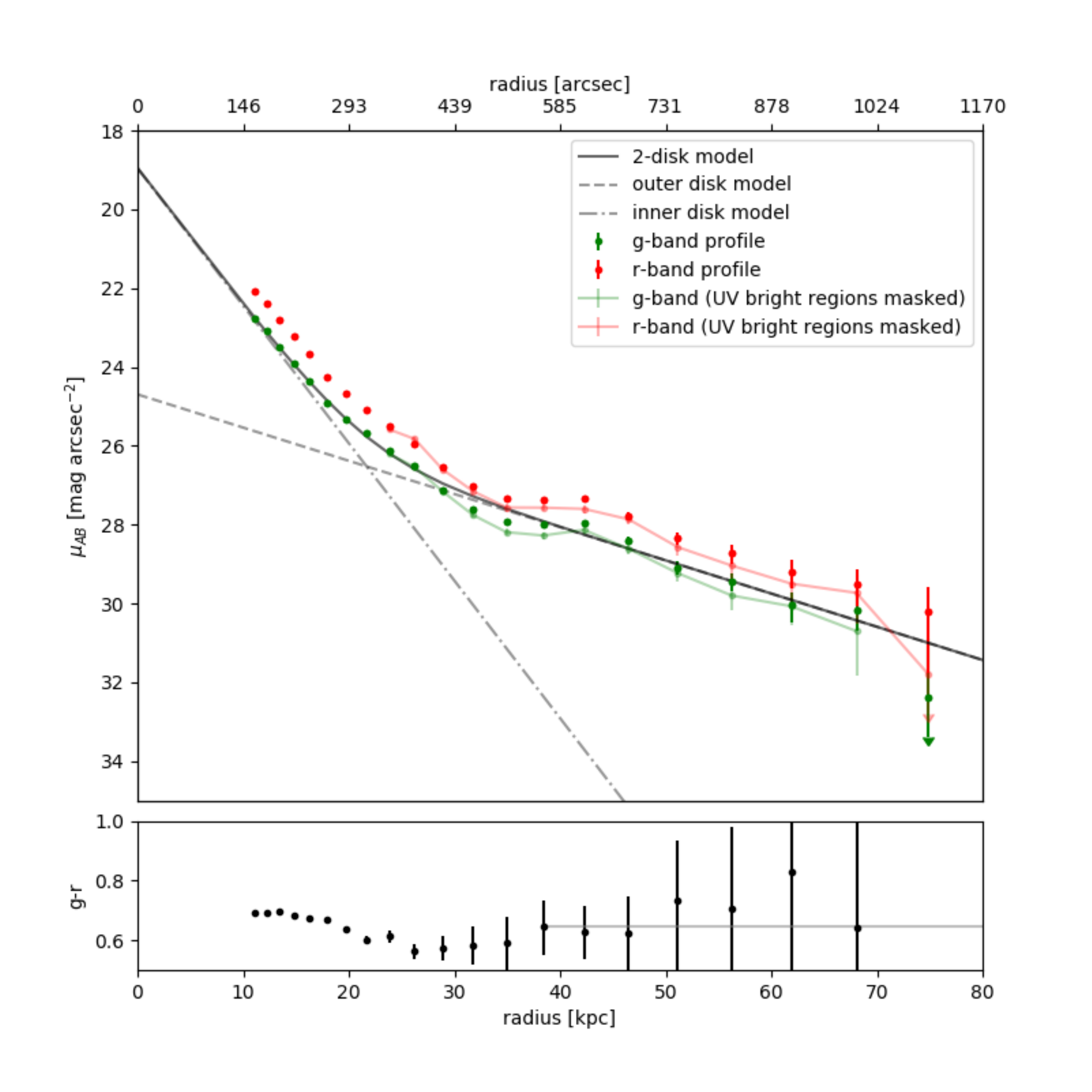}
\caption{Top: Surface brightness profile of NGC 2841 in Sloan \textit{g} and \textit{r}-band. A 2-disk model is fit to the surface brightness profile. The inner and outer disks have disk scale lengths of 3.1$\pm$0.1 kpc and 13$\pm$3 kpc respectively. Plotted in a lighter shade of green and red with lines connecting the data points is the derived surface brightness profile after masking the UV bright regions. Bottom: The (\textit{g}-\textit{r}) color profile. The error bars in both plots include RMS and sky errors. } 
\label{fig:profile}
\end{figure}

In order to make \textit{g} and \textit{r}-band surface brightness profiles, stars and other sources were masked so that they did not contribute to the low surface brightness outer disk. The method used to create the mask for making the surface brightness profile was similar to that for measuring the sky level. The only difference is that NGC 2841 was removed from the segmentation map used to create the mask and the step to mask NGC 2841 by visual inspection in SAOImage DS9 was not applied. The mask for the bright star to the east of the galaxy (see Figure~\ref{fig:image}) overlaps with the central bulge of NGC 2841. Based on previous surface brightness profiles created of the inner disk, the bulge is no longer dominant beyond 60 arcseconds~\citep{Boroson1981}, so we can safely fit a pure exponential disk to the surface brightness profile beyond 200 arcseconds, and this is where the surface brightness profile in this paper begins. To create the profile, isophotes were fitted using the {\tt iraf.stsdas.isophote.ellipse} routine~\citep{Jedrzejewski1987} in PyRAF\footnote{PyRAF are products of the Space Telescope Science Institute, which is operated by AURA for NASA}. An initial x and y coordinate for the center of the galaxy was input, but the center, position angle and ellipticity of isophotes were allowed to vary. However, beyond a radius of $\sim$35 kpc, there is no longer enough signal to noise for the ellipse-fitting routine to allow these parameters to vary, and so the ellipse shape is fixed beyond that radius. The routine also extracts the average unmasked pixel value in each isophote. The \textit{g} and \textit{r}-band surface brightness profiles of NGC 2841 are shown in Figure~\ref{fig:profile}, where the error bars include random as well as systematic sky-errors.   

The visible light in the outskirts of NGC 2841 is clearly part of an extended disk, because the visible wavelength morphology of the galaxy traces the HI and UV disks. A 21-cm kinematic study of the galaxy shows there is a warp in the gas disk~\citep{Bosma1978}. Visually, \textit{g} and \textit{r}-band light beyond $\sim$30 kpc is dominated by a warped outer disk, aligned with the HI and the UV star forming disk. The warp is most obvious in the \textit{g}-\textit{r} color, and the HI images. As a further indicator of a stellar disk warp, the position angle of the fitted elliptical isophote jumps from -30 $\pm$1.5 degrees clockwise from the y-axis within 30 kpc to -26.5 at 30 kpc. Note, however, that the position angle is not allowed to vary in the ellipse fitting routine beyond $\sim$35 kpc due to low signal to noise. Because the disk is warped, and because the surface brightness profile appears to up-bend, a two-exponential disk model is most appropriate and this was fit to the galaxy surface brightness profile. The inner and outer disks have scale lengths of 3.1$\pm$0.1 kpc and 13$\pm$3 kpc, respectively. The two-disk model is shown in Figure~\ref{fig:profile}, with the solid grey line being the sum of the two components. Our measured surface brightness profile extends to $\sim$70 kpc, which is $\sim$23 inner disk scale lengths. 

\subsection{Is the outer disk light contaminated by scatter from the wide-angle PSF?}

As described in the Introduction, the wide-angle point-spread function can play an important role in the measurement of profiles at very low surface brightness levels. Since the visibility of the stellar aureole varies as a function of atmospheric conditions, part of the data reduction pipeline for Dragonfly data rejects science exposures with zeropoints that deviate from a nominal zeropoint by more than $\sim$0.1 mag. This procedure removes the science exposures most contaminated by scattered light from the PSF. To test the significance of remaining contamination, we convolved a measured PSF with a one-dimensional model galaxy profile similar to NGC 2841 to observe the change in the surface brightness profile at large radii. A bulge central surface brightness of $\mu_0=20.1$ and bulge effective radius of $R_e=0.94$ kpc was used~\citep{Boroson1981} together with the two-disk model found in this paper. 

The measured PSF has a radius of 10 arcmin and spans 18 magnitudes in surface brightness. The inner part of the PSF was measured using the brightest unsaturated star in the field. The outer part of the PSF was measured using the brightest saturated star in the field. The IRAF\footnote{IRAF is distributed by the National Optical Astronomy Observatory, which is operated by the Association of Universities for Research in Astronomy (AURA) under a cooperative agreement with the National Science Foundation.} routine {\tt pradprof} was used to compute a radial profile around each star, which was then median binned in the radial direction to remove contamination by other sources. We note that this can only overestimate the PSF compared to the PSF that would be obtained without contamination from other sources. 

The outcome of this exercise was that, because of the careful control of systematics in our experimental setup, the surface brightness profile of NGC 2841 remains unaffected by the wide-angle PSF down to at least $\mu=32$ mag arcsec$^{-2}$. 

\subsection{How is the extended light distributed?}

Is the extended visible wavelength emission from NGC 2841 simply the visible wavelength counterpart of the UV knots identified by GALEX? Or is this light truly distributed at all azimuthal angles around the disk? 

To explore whether the visible wavelength light in the outer regions of NGC 2841 is entirely the visible wavelength emission from the UV knots identified by GALEX, we measured the \textit{g} and \textit{r}-band surface brightness profiles again after masking out the UV bright regions to see how much signal is left outside of the star forming regions. This surface brightness profile is shown also shown in Figure~\ref{fig:profile} as line-connected green and red data points for the \textit{g} and \textit{r}-band profiles, respectively. While the surface brightness profiles in both \textit{g} and \textit{r}-band have dimmed (by $\sim20\%$ beyond 30 kpc) as a result of this masking, the overall shape and extent of the light profiles remain similar. We therefore conclude that the outer disk light is not simply the visible wavelength counterpart of the UV knots identified by GALEX. 

\begin{figure}[!htbp]
\centering
\includegraphics[width=0.52\textwidth]{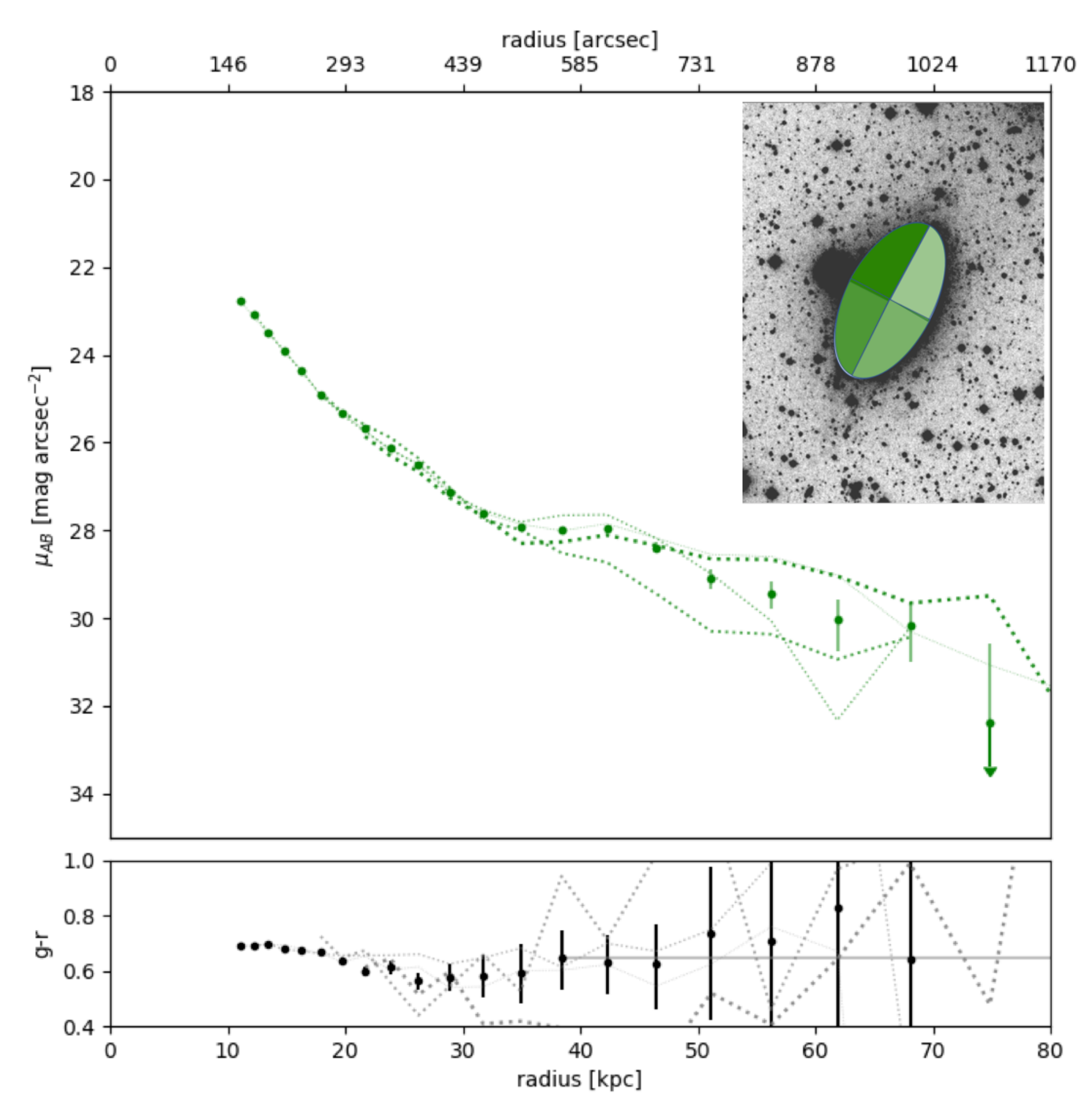}
\caption{Top: The surface brightness profile of NGC 2841 in the Sloan \textit{g} filter is shown again in green data points with error bars. Dotted green line profiles are azimuthal wedge \textit{g}-band profiles. The darkest dotted line corresponds to the darkest green wedge shown in the top right-hand side image, with subsequent lighter shades of green corresponding to other lighter green wedges. Bottom: The (\textit{g}-\textit{r}) color profile is shown again in black. Dotted grey lines are the azimuthal wedge color profiles. The darkest grey dotted line corresponds to the darkest green wedge shown in the top right-hand side image, with subsequent lighter shades of grey corresponding to other lighter green wedges. The error bars in both plots include RMS and sky errors and are indicative of error bar size for all azimuthal wedge profiles.} 
\label{fig:profile_wedge}
\end{figure}

A surface brightness profile averages light from all angles. This means in the low surface brightness galaxy outskirts the profile shape might be dominated by features in a small azimuthal wedge of the galaxy. To see if there is extended galaxy emission at all azimuthal angles, surface brightness profiles for azimuthal wedges were measured and plotted in Figure~\ref{fig:profile_wedge}. While there is scatter in the profile in different azimuthal wedges, there is consistently light at all angles, lending further evidence to a smooth underlying disk that is the continuation of disk visible in Figure~\ref{fig:image}.

\subsection{Mass of stellar disk and comparison to the gas disk}
\begin{figure}[!htbp]
\centering
\includegraphics[width=0.5\textwidth]{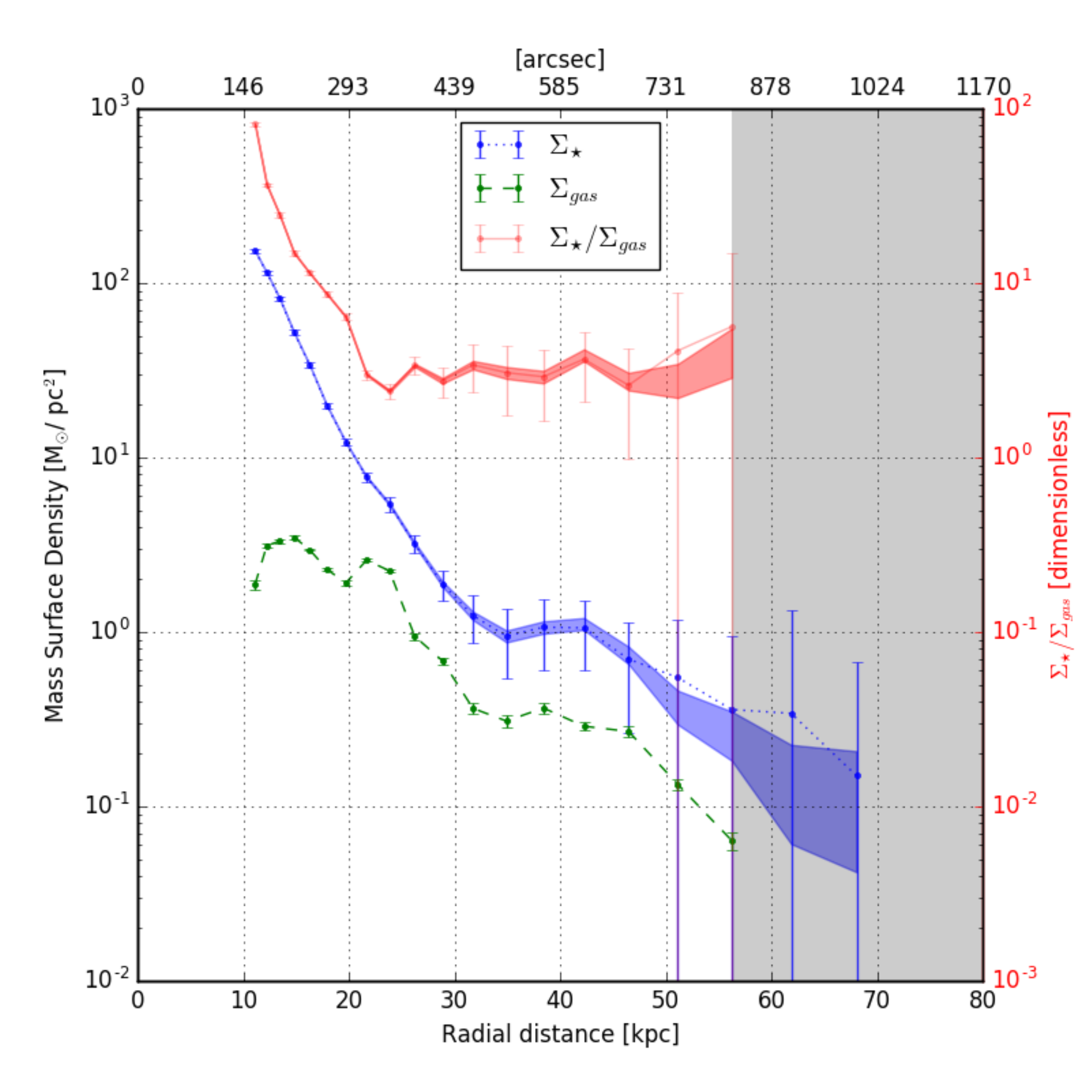}
\caption{The mass surface density of stars and gas ($\Sigma_{\star},\Sigma_{gas}$ $[\text{M}_{\odot} \text{ pc}^{-2}]$) and the ratio of the two ($\Sigma_{\star}/\Sigma_{gas}$) are plotted as a function of radius for NGC 2841. The error bars include the RMS uncertainty and the systematic error due to sky uncertainty in the g-band image and the (\textit{g}-\textit{r}) color used to calculate the stellar mass surface density. The shaded profiles assume that the color in the outer disk beyond 40 kpc is a constant. See the text for more details.}  
\label{fig:profile_masses}
\end{figure}

The stellar and the gas mass surface density profiles of NGC 2841 are shown in Figure~\ref{fig:profile_masses}. The stellar mass surface density was calculated in the same way as described in~\cite{vanDokkum2014}, using relations given in~\cite{Bell2001}:
\begin{eqnarray}
\log_{10}(\rho [\text{M pc}^{-2}])= &-& 0.4(\mu_g [\text{mag arcsec}^{-2}] - \text{DM}) \nonumber \\
                                    &+& 1.49(g-r) \nonumber \\
                                    &+& 1.64 + \log_{10}(1/C^2)
\label{eqn:stellarmass}
\end{eqnarray}
Using a distance of 14.1 Mpc~\citep{Leroy2008}, $\text{DM}=30.7$ is the distance modulus and $\text{C}=0.0684$ is the conversion factor from arcsecond to kiloparsecs. 

Since the mass density depends on color, we explored the impact of color on the inferred mass density using two different approaches. Densities obtained using both approaches are shown in Figure~\ref{fig:profile_masses}, as error bars and as shaded regions. In the first approach, we used the measured (\textit{g}-\textit{r}) color, and its uncertainty, which includes RMS and sky errors in both filters. The uncertainties in the mass measurements using this approach are displayed as error bars and include color and \textit{g}-band RMS and systematic sky errors. In the outskirts, sky errors dominate. This is the most conservative indication of how well the stellar mass in the outskirts of NGC 2841 can be measured. The second approach, which is plotted as the shaded region in Figure~\ref{fig:profile_masses}, was to use the measured (\textit{g}-\textit{r}) color within 40 kpc, and a constant value of 0.65 beyond that. This choice of color is indicated by the grey horizontal line in the lower panel of Figure~\ref{fig:profile}. Our rationale for adopting this constant color is because beyond 50 kpc the measurements of (\textit{g}-\textit{r}) color have very large uncertainties due to uncertainties in the sky level in both \textit{g} and \textit{r}-band. One can view the shaded region as a potential stellar mass surface density profile if the color in the outer disk remains at a constant 0.65 beyond 40 kpc. The errors indicated by the shaded region in the stellar mass surface density in Figure~\ref{fig:profile_masses} include both RMS errors and systematic sky-errors in the \textit{g}-band data.

The gas mass surface density was calculated using equation A1 from~\cite{Leroy2008}:
\begin{equation}
\Sigma_{\text{gas}} [\text{M}_\odot \text{pc}^{−2}] = 0.020 \cos i \text{ } I_{21 \text{cm }}[\text{K km s}^{-1}]
\label{eqn:gasmass}
\end{equation}
where $i=1.29$ is the inclination in radians~\citep{Leroy2008}. $I_{21 \text{cm}}$ is the 21 cm flux from the THINGS HI map. This gas mass equation includes a factor 1.36 to reflect the presence of helium. The gas mass surface density is plotted in green in Figure~\ref{fig:profile_masses}. Error bars shown account for the RMS scatter in each isophotal annulus. The shaded grey region is where the THINGS HI map and the GALEX FUV map has no detection at their respective sensitivity limits.

To better illustrate the relationship between stellar and gas mass, the ratio of the two is plotted in Figure~\ref{fig:profile_masses} on the right-hand axis, in red. The ratio of stellar mass to gas mass remains remarkably constant (at 3:1) from just beyond $\sim$20 kpc, to the limit of the THINGS data. Remarkably, at the sensitivity limit of the THINGS survey and at the current depth of Dragonfly's observations of NGC 2841, there is no radius at which the mass surface density of HI gas begins to dominate over that of the stars. This has been measured out to 50 kpc, or 16 inner disk scale lengths. At radii greater than 50 kpc, the uncertainty in the sky levels of both the \textit{g} and \textit{r} Dragonfly images means the stellar mass is not measured with enough precision to conclude it is greater than the gas mass. However, the color in the disk beyond 50 kpc would have to be bluer than (\textit{g}-\textit{r}) = 0.3 in order for the stellar mass to drop below that of the gas mass. 

\subsection{Timescales}

\begin{figure}[!htbp]
\centering
\includegraphics[width=0.5\textwidth]{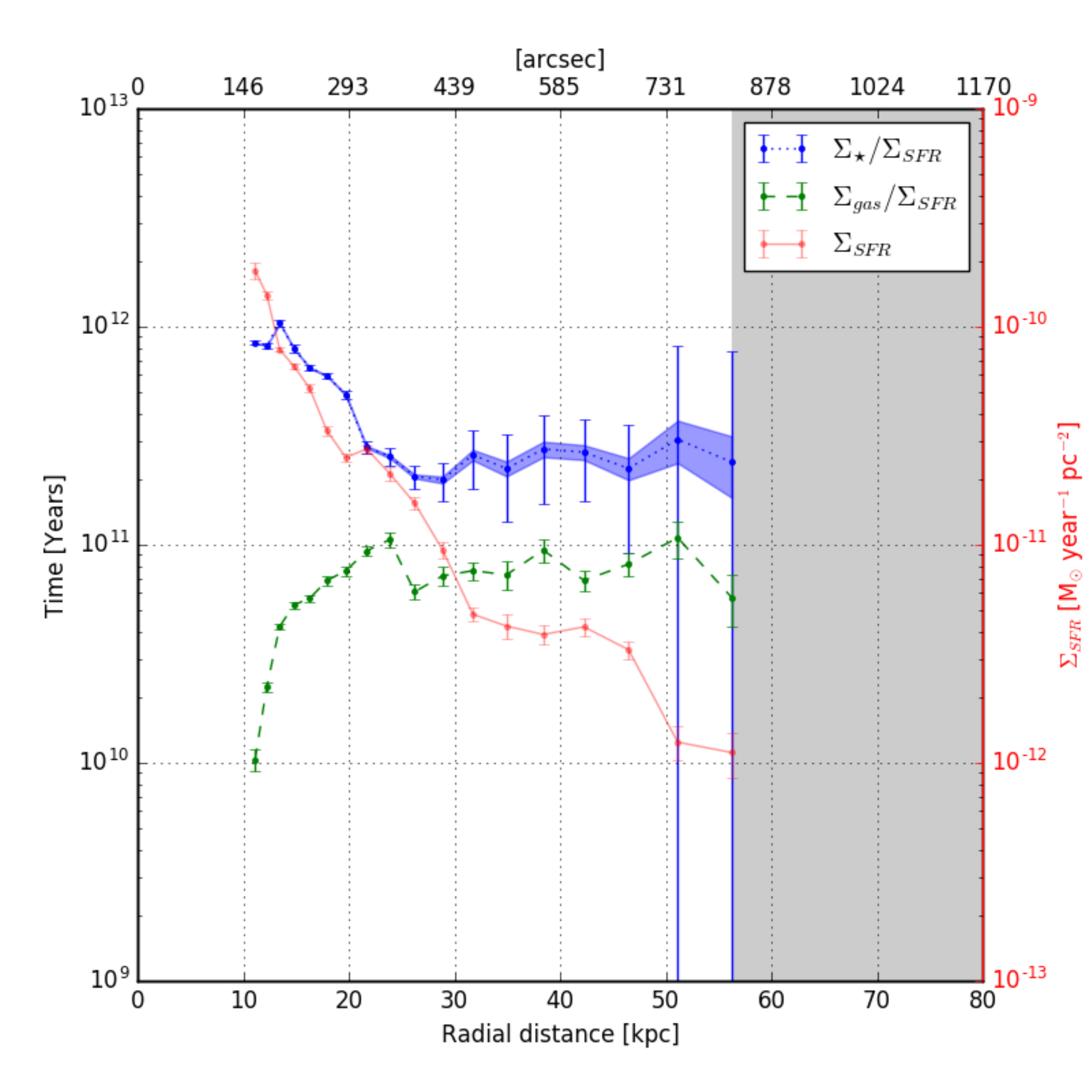}
\caption{The stellar mass buildup time ($\Sigma_{\star} / \Sigma_{\text{SFR}}$, blue) and gas depletion time ($\Sigma_{\text{gas}} / \Sigma_{\text{SFR}}$, green) and SFR surface density (black) are shown as a function of radius. The error bars include the RMS uncertainty in the HI, UV, \textit{g}-band and \textit{r}-band measurements, as well as the systematic error due to sky uncertainty in the g-band image and the (\textit{g}-\textit{r}) color used to calculate the stellar mass surface density. }
\label{fig:profile_timescales}
\end{figure}

The gas depletion time ($\Sigma_{\text{gas}} / \Sigma_{\text{SFR}}$) and the stellar mass buildup time ($\Sigma_{\star} / \Sigma_{\text{SFR}}$) for NGC 2841 are plotted as a function of radius in Figure~\ref{fig:profile_timescales}. The star formation rate (SFR) surface density used was obtained from the GALEX FUV image using equation 1 taken from~\cite{Wilkins2012}:
\begin{eqnarray}
&\text{ }&\Sigma_{\text{SFR}} [\text{M}_{\odot} \text{ year}^{-1} \text{ pc}^{-2}] \nonumber \\
&=& 10^{-34} \text{ L}_{\text{FUV}} [\text{ergs} \text{ s}^{-1} \text{ Hz}_{-1} \text{ kpc}^{-2}] \text{ B}_{\text{FUV}} 
\label{eqn:SFR}
\end{eqnarray}
where $\text{B}_{\text{FUV}}$ varies with the slope of the initial mass function (IMF). The value used here is $0.9 \pm 0.18$, based on the Kennicutt 1983 IMF~\citep{Kennicutt1983}. The SFR surface density is also plotted in Figure~\ref{fig:profile_timescales}, on the right-hand axis, in red. The SFR and HI mass surface density error bars only includes the RMS scatter in each isophotal annulus. The error bars in  $\Sigma_{\star} / \Sigma_{\text{SFR}}$ include fractional errors in $\Sigma_{\star}$ and $\Sigma_{\text{SFR}}$ added in quadrature. For $\Sigma_{\star}$, the error bars were calculated using two methods in the same way as in Figure~\ref{fig:profile_masses}: the error bars include the \textit{g} and (\textit{g}-\textit{r}) errors while the shaded profile uses a (\textit{g}-\textit{r}) color model and only include the \textit{g}-band errors.

The stellar mass buildup and gas depletion timescales remain constant from beyond $\sim$20 kpc, the same region with a constant 3:1 ratio of stellar to gas mass surface density. The stellar mass buildup time in this region is ~250 Gyr and the gas depletion time is on the order of 70 Gyr. Both of these timescales are much longer than the age of the Universe. 

\section{Discussion}
\label{sec:discussion}

\subsection{The origin of outer disk stars}

The underlying stellar disk in NGC 2841 discovered using Dragonfly is gigantic, reaching beyond the size of the most sensitive HI and UV disk observations to $\sim$70 kpc ($\sim$5 $R_{25}$ or $\sim$23 inner disk scale lengths). The surface brightness profile of the galaxy shows an upward bending break at 30 kpc, when the \textit{g}-band surface brightness is $\sim$28 mag arcsec$^{-2}$. Similar upward bending (Type I) surface brightness profiles are common at large radii, measured using a combination of photometry and star counts~\citep{Barker2009,Barker2012,Watkins2016}. Of note is a multi-object spectroscopy study of stars in M31 by ~\cite{Ibata2005}, which traced disk stars out to $\sim$70 kpc. In NGC 2841, the position of the upbend corresponds to the start of a low surface brightness warp in the outer disk. This warp is visible in both \textit{g} and \textit{r}-band, THINGS HI and GALEX UV images. One common assumption is that in the outer disks of galaxies, neutral gas is the dominant baryonic component based on the observation that in general HI disks extend much further than stellar disks~\citep{vanderKruitFreeman2011,Elmegreen2016}. A comparison of the stellar mass to gas mass surface densities shows that for NGC 2841, there is no radius at which the mass surface density of gas begins to dominate over that of the stars. Beyond $\sim$20 kpc, NGC 2841 also has the interesting property that the stellar to gas mass surface density ratio is a constant 3:1.

A central question is: how did this giant stellar disk form? There are three main ways to populate the outer disk with stars: (1) stellar migration, (2) accretion of stars, and (3) in-situ star formation. Note that one mechanism does not preclude the others. We discuss the merits and weaknesses of each of these possibilities below.

(1) \textit{Stellar Migration:} \cite{SellwoodBinney2002} showed that transient spiral arms in galaxy disks can scatter stars into orbits at different radii, but allow the stellar orbits to remain circular. Subsequent simulations by Ro{\v s}kar~\citep{Roskar2008a, Roskar2008b} showed that a downward bending (type II) surface brightness profile can be explained by a star formation threshold combined with stellar migration, which can move stars beyond their formation radius. There are several issues with appealing to stellar migration to populate the outer disk of NGC 2841. Firstly, it is unclear whether stellar migration can move so much mass to such a large range of radii beyond 30 kpc.~\cite{Watkins2016} has similar concerns with appealing to stellar migration as a means of populating the outer disk of three nearby galaxies, where stars have to be moved several disk scale lengths beyond the extent of spiral arms. Secondly, stellar migration should not create an upwards bending surface brightness profile. Thirdly, the stars, as well as the gas, beyond 30 kpc are in a warped disk, and it is unclear how the migrated stars would end up in such a warped orbit. Perhaps the stars were displaced to large radius long ago and the warp was induced by a later interaction that warped both the stars and the gas.

(2) \textit{Accretion:} In order for accreted stars to build up a co-planar disk, incoming stars need to have a narrow range in angular momenta that matches the existing disk, otherwise simulations show that the accreted stars tend to end up in a bulge or stellar halo~\citep{Toomre1977,Schweizer1990}. In between these extremes lies infall with a slight mismatch in angular momentum, the result of which is a warped disk~\citep{Binney1992}. The warp in the disk of NGC 2841 beyond 30 kpc may hint at past accretion onto the disk with a slightly different angular momentum than that of the underlying stellar population. A quarter of a century ago,~\cite{Binney1992} noted presciently that: ``...it is by no means inconceivable that warps are in direct physical contact with material that is only now joining the galactic system''. Since our data for NGC241 extends out to $\sim$60 kpc, it is tempting to associate the outer warped disk in this system with material that has only recently infallen. If this interpretation is correct, we have reached far enough into the outskirts of NGC 2841 to be probing its circumgalactic environment, and may be witnessing a slightly more evolved part of the cool-flow driven galactic component referred to by ~\cite{Bland-Hawthorn2017} as the `proto-disk'. 

The total stellar mass in the disk beyond $\sim$30 kpc (where the warped disk starts to dominate) is $1.4\pm0.2\times10^8\text{ M}_{\odot}$. For comparison, the Small Magellanic Cloud has a stellar mass of $\sim$7$\times10^8\text{ M}_{\odot}$~\citep{Dooley2017}. The outer disk could potentially be formed by one or several small accretion events with the angular momenta of incoming dwarf galaxies almost aligned with that of NGC 2841. 

The similarity in angular momenta required in this scenario may not be particularly improbable, since it seems that satellites of nearby galaxies can map out organized structures. Well-known examples include the Great Plane of Satellites around M31~\citep{Ibata2013} and the Vast Polar Orbital structure around the Milky Way Galaxy~\citep{Pawlowski2015}. One hypothesis for the existence of these planar structures is accretion along large scale filamentary structures~\citep{Ibata2013}. If this is the case, the accreted dwarf galaxies could deposit their gas and stars onto the outer disk, possibly creating a constant stellar mass to gas mass ratio after a few rotations.  

(3) \textit{In-Situ Star Formation:} Stars in the outer disk of NGC 2841 clearly trace the distribution of the HI gas. This observation tends to favor a model in which the stars were formed in-situ. However, at the current SFR, it would take $\sim$200 Gyr to build up the outer disk stellar mass. On the other hand, the global star formation in the Universe was much higher in the past than today, peaking at redshifts of 2-3, with a SFR an order of magnitude greater than today~\citep{Madau2014}. At a star-formation rate that is 10 times that of the current SFR in NGC 2841, it would take $\sim$15 Gyr to build up the outer disk stellar mass. This is only slightly longer than a Hubble time, so such a scenario could be made to work. If in-situ star formation is responsible, then it begs the question of how the outer disk achieved such high levels of star formation in the past. The entire disk is currently Toomre stable and star formation is occurring in UV knots. It is possible the gas density in the outer disk was higher in the past, but this may not be enough to stimulate sufficient star formation. \cite{Cormier2016} compared a sample of HI-rich galaxies to a control sample and showed that there is no increase in molecular gas mass or SFR in the outer disk in the HI rich sample. The difference was that the HI rich sample was able to able to sustain star formation in the outer disk for a longer period of time. Other studies support this by showing that gas depletion takes longer than a Hubble time and that SFRs and the Toomre stability parameter are not correlated~\citep{Bigiel2010}. \cite{Semenov2017} carried out simulations that suggest global gas depletion times are long because only a small fraction of gas is converted into stars before star-forming regions are disrupted. Therefore, gas has to cycle in and out of the star-forming state many times before being turned into stars.

None of the mechanisms described above strike us as unreasonable, so perhaps the most likely scenario is that the outer disk of NGC 2841 is being built up by a combination of multiple mechanisms. For example, cool flow infall of gas which somehow triggers in-situ star-formation. However, one then wonders how these mechanisms conspired together to result in a constant ratio between stellar and gas mass surface density beyond $\sim$20 kpc. 

The present paper reports results for a single galaxy, NGC 2841, the results from which indicate that with experimental setups optimized for low surface brightness imaging, stellar disks can be probed out to radii where the disk starts to warp. This may be an indication we are seeing parts of the disk that encroach upon the circumgalactic medium. Future papers will carry out similar analyses on other galaxies in the Dragonfly Nearby Galaxies survey, four of which have accompanying THINGS HI data. If NGC 2841 is any guide, the key questions for understanding galactic outskirts must now include: what fraction of massive spirals contain enormous underlying stellar disks? Are these disks always more massive than gaseous disks revealed by HI imaging? Is the mass ratio of stars to gas a constant in the outer disk, as seen in NGC 2841? At ultra-low surface brightness levels, do stellar disks always trace HI, and are these disks always warped in a manner consistent with infall? Is the geometry of the warps correlated with the positions of companion galaxies, as would be expected if the warped disk is built up by infall, and companion galaxies trace dark matter filaments? This long list of questions befits the richness of the phenomena being revealed at low surface brightness levels in the outskirts of galaxies. In any case, it seems to us that the key to answering these questions is to approach them in the appropriate panchromatic context, focusing on comparisons of surface mass densities, and not just on arbitrary definitions of the `sizes' of disks at various wavelengths.

\section{Acknowledgements}

We thank the anonymous referee for a thoughtful and constructive report, which improved the paper. Support from NSERC, NSF grants AST-1312376 and AST-1613582, and from the Dunlap Institute (funded by the David Dunlap Family) is gratefully acknowledged. We thank the staff at New Mexico Skies Observatory for their dedication and support. JZ thanks Rhea-Silvia Remus for useful discussions. 
Funding for the Sloan Digital Sky Survey IV has been provided by the Alfred P. Sloan Foundation, the U.S. Department of Energy Office of Science, and the Participating Institutions. SDSS-IV acknowledges support and resources from the Center for High-Performance Computing at the University of Utah. The SDSS web site is www.sdss.org.

\bibliographystyle{aasjournal}

\clearpage

\end{document}